**Title:** Personalized Mathematical Model Predicting Endotoxin-Induced Inflammatory Responses in Young Men

**Authors:** Renee Brady[1], Dennis O. Frank-Ito[2], Hien T. Tran[1], Susanne Janum[3], Kirsten Møller[4], Susanne Brix[5], Johnny T. Ottesen[6], Jesper Mehlsen[3], and Mette S. Olufsen[1]

[1]*Department of Mathematics, North Carolina State University, Raleigh, NC,*
[2]*Departments of Surgery, Duke University Medical Center, Durham, NC,*
[3]*Coordinating Research Centre, Frederiksberg Hospital, Frederiksberg, Denmark*
[4]*Department of Neurointensive Care, Rigshospitalet, University of Copenhagen, Denmark*
[5]*Department of Systems Biology, Technical University of Denmark, Lyngby, Denmark*
[6]*Department of Science and Environment, Roskilde University, Roskilde, Denmark*

*Mette S. Olufsen, NCSU Department of Mathematics, Box 8205, Raleigh, NC 27695, Email: msolufse@ncsu.edu, Tel: (919) 515-2678*

**Sources of support**

Olufsen and Brady were supported in part by NSF-DMS Award 1246991 and Olufsen was supported in part by NIH-NIGMS subaward to NCSU 1P50GM094503-01A0. Møller was responsible for experimental studies supported by Grosserer Ehrenreichs Foundation, Aase and Ejnar Danielsens Foundation, The Danish Society of Anaesthesiology and Intensive Care Medicine (DASAIM), The Medical Society of Copenhagen, and King Christian the X's Foundation.

**Running Head:** Personalized Inflammation Model for Young Men

ABSTRACT – The initial reaction of the body to pathogenic microbial infection or severe tissue trauma is an acute inflammatory response. The magnitude of such a response is of critical importance, since an uncontrolled response can cause further tissue damage, sepsis, and ultimately death, while an insufficient response can result in inadequate clearance of pathogens. A normal inflammatory response helps to annihilate threats posed by microbial pathogenic ligands, such as endotoxins, and thus, restore the body to a healthy state. Using a personalized mathematical model, comprehension and a detailed description of the interactions between pro- and anti-inflammatory cytokines can provide important insight in the evaluation of a patient with sepsis or a susceptible patient in surgery. Our model is calibrated to experimental data obtained from experiments measuring pro-inflammatory cytokines (interleukin-6 (IL-6), tumor necrosis factor (TNF-$\alpha$), and chemokine ligand-8 (CXCL8)) and the anti-inflammatory cytokine interleukin-10 (IL-10) over 8 hours in 20 healthy young male subjects, given a low dose intravenous injection of lipopolysaccharide (LPS), resulting in endotoxin-stimulated inflammation. Through the calibration process, we created a personalized mathematical model that can accurately determine individual differences between subjects, as well as identify those who showed an abnormal response.

# INTRODUCTION

Invasion by disease or injury triggers an acute inflammatory response that is vital in the repulsion of the pathogens and the induction of a repair mechanism in damaged tissues. A typical inflammatory response consists of the following: 1) phagocytic cells are activated, 2) pro- and anti-inflammatory mediators are triggered, 3) the invading pathogen is cleared, 4) the tissue is repaired if necessary, and 5) the response is subdued. An insufficient response can lead to persistent tissue injury, resulting in conditions such as autoimmune diseases, cancer, and lifestyle-related disorders [1]. An uncontrolled, excessive production of pro-inflammatory cytokines from immune cells and traumatic tissues can cause systemic inflammatory response syndromes such as sepsis and, in life-threatening cases, septic shock [2]. The Agency for Healthcare Research and Quality lists sepsis as the most expensive condition treated in U.S. hospitals, costing more than $20 billion in 2011 [3]. Toll-like receptor-4 (TLR4) signaling, responsible for the production of the inflammatory mediators, may be a key pathway in the pathophysiology of sepsis. Thus, understanding TLR4 signaling and the mediators produced is critical in evaluating patients experiencing sepsis and those undergoing surgery, such as knee or hip replacement, who may be more susceptible to sepsis.

Experimentalists have studied the inflammatory response in mice and humans through the administration of specific pathogens, particularly through endotoxin, a cell wall component of gram-negative bacteria. In mice, these studies have provided great insight into the inflammatory response. However, due to physiological differences between mice and human TLR4 activation [4], a key pathway in the pathophysiology of sepsis, as well as a human's sensitivity to the effects of endotoxin, similar strides have not been made in understanding the acute inflammatory response in humans. To quantify the differences in the inflammatory responses between mice and humans, Copeland et al. [5] conducted an experiment in which mice and humans were given equivalent doses of endotoxin and the levels of circulating cytokines TNF-α and IL-6 were measured and compared. The study found that humans experienced a rapid physiological response, consisting of fever, tachycardia, and slight hypotension, which was not evident in mice. Thus, it was concluded that the autonomic control system is affected by the inflammatory response in humans, but likely not in mice.

Generalized mathematical models of the acute inflammatory response in humans propose that the response to endotoxin consists of an instigator and a set of pro- and anti-inflammatory mediators working in unison to restore homeostasis [6, 7]. These models are formulated as a system of ordinary differential equations set up to integrate known biological assumptions. Simple models have the advantage that they allow rigorous mathematical analysis and use simplified biological assumptions. Higher-order models [8, 9] have been developed to predict the generalized inflammatory response in mice. These include biological complexity predicting the dynamics of individual cytokines. However, these models are too complex, both conceptually and computationally due to the inherent nonlinearity and the large number of unknown inputs to the model to analyze mathematically.

Our study was motivated by the higher-order model proposed by Clermont et. al. [9] predicting the inflammatory response in mice. The model includes neutrophils and macrophages directly activated by bacterial endotoxin (*E. coli* lipopolysaccharide (LPS)) and indirectly via systemic stimuli produced by trauma and hemorrhage. The activated phagocytic cells promote the production of pro-inflammatory cytokines such as tumor necrosis factor (TNF-$\alpha$) and interleukins 6 and 12 (IL-6, IL-12) and anti-inflammatory cytokines such as IL-10. This model was used to reproduce qualitative results from three separate scenarios in mice (trauma, surgical trauma/hemorrhage, and 3 and 6 mg/kg of endotoxin). Expanding upon these results, Chow et al. [8] developed a 15-state model of the acute inflammatory response in mice to endotoxin, hemorrhage, and surgical trauma. Mice were administered endotoxin at levels of 3, 6, or 12 mg/kg and experimental data was collected—including TNF-α, IL-6, IL-10 and nitric oxide byproducts. The insertion of a cannula induced surgical trauma, and hemorrhagic shock was stimulated by blood withdrawal. The mathematical model was calibrated to experimental data from each of these scenarios.

Although the model proposed by Chow et al. [8] provided insightful information about modeling the dynamics of pro- and anti-inflammatory mediators, their model was designed to capture this phenomenon in mice. Furthermore, the majority of mathematical modeling and analyses on endotoxin-induced signaling and cytokine production in monocytes and macrophages have been done in mice because of the ability to calibrate these models to *in vivo* experimental data. On the contrary, such experiments cannot be easily done in humans since endotoxin challenges are potent immunostimulators and are highly regulated in human studies, especially in the US.

In this study, we have developed a personalized mathematical model of the acute inflammatory response to endotoxin challenge, based on the biology in humans and the interactions between pro- and anti-inflammatory cytokines. This model has the potential to advance current understanding in evaluating patients during the early stages of recovery after surgery, when many are encouraged to regain mobilization as soon as possible [10]. Our model is calibrated to experimental data from 20 healthy young men who were administered an intravenous (i.v.) endotoxin dose of 2 ng/kg of body weight. Concentrations of the cytokines IL-6, CXCL8, TNF-$\alpha$, and IL-10 were measured hourly for 8 hours. Our model can accurately predict the dynamics of these cytokines up to 8 hours after the introduction of the inflammatory agent on an individual basis, including those individuals who exhibit an abnormal response.

## MATERIALS AND METHODS

*Experimental Data*

*Study Participants.* Twenty healthy, young male volunteers, between the age of 20 and 33 years (median 24.3 years), were recruited via public advertising from the general population in Copenhagen, Denmark to participate in this study. The inclusion criteria were as follows: (i) male, age 18-35, (ii) good general health, demonstrated by medical history and medical examination, (iii) Body Mass Index (BMI) $<$ 30 kg/m$^2$ and (iv) written informed consent prior to enrollment. The exclusion criteria were as follows: (i) daily medicine intake (excluding antihistamines during pollen season), (ii) smoking or use of nicotine substitutes, (iii) previous allergic reaction to nicotine pads, and (iv) previous splenectomy. The study protocol was approved by the Regional Committee on Health Research Ethics (protocol-ID H-3-2012-011) and the Regional Data Monitoring board (ID j-2007-58-0015, local 30-0766) and reported to clinicaltrials.gov (NCT01592526).

*Experimental Procedure/Design.* This was an open-label, randomized cross-over study in which participants received a bolus of endotoxin at a dose of 2 ng/kg of body weight 2 hours after the start of the experiment. Blood samples were collected before the endotoxin infusion (at $t = 0$) and then at $t = 2, 3, 3.5, 4$ hours and in one hour increments for a total of 8 hours.

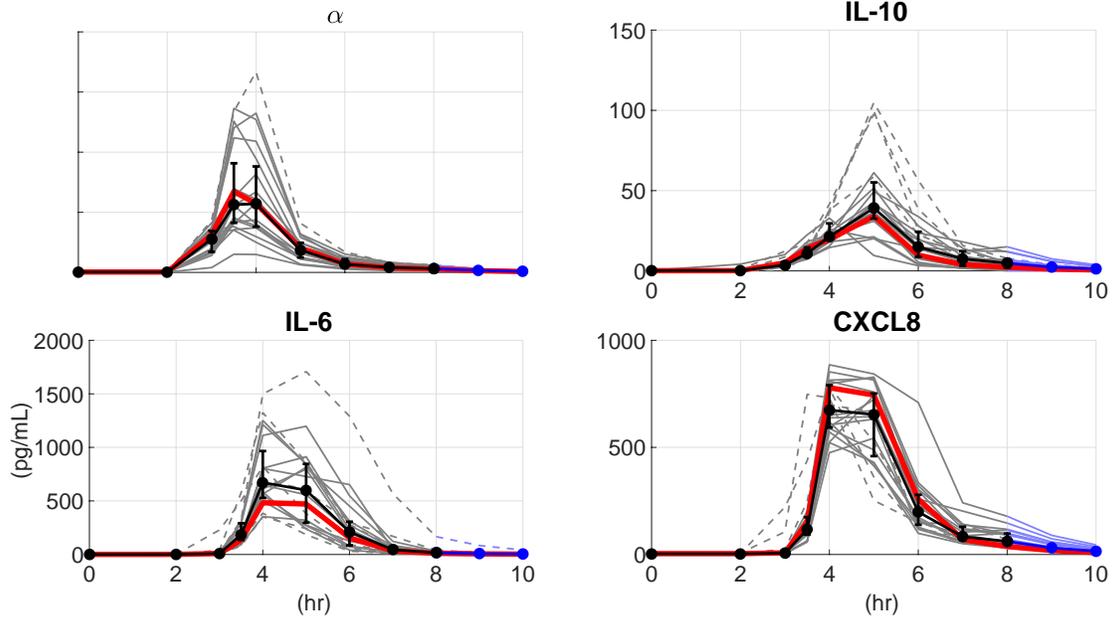

FIG 1: **Experimental Data.** Plasma cytokine responses to intravenous endotoxin administration in 20 healthy young men. Median (black circle), interquartile range (error bars), and subject most in line with data mean (red) are depicted. Abnormal response (identified via Box-and-Whisker plots) are denoted by dashed lines. Pro-inflammatory cytokines, TNF-$\alpha$, IL-6, and CXCL8, and the anti-inflammatory cytokine IL-10 levels were measured at t = 0, 2, 3, 3.5, 4 h and in one hour increments for the next 4 hours. Pseudodata was added at $t = 9$ and 10 to ensure that cytokines decayed to baseline levels (blue). Endotoxin was administered at t = 2.

*Blood Collection.* Blood samples for the analysis of cytokine levels in plasma were collected in EDTA tubes (Greiner bio-one, Germany). The samples were kept on ice until centrifuged at 4 °C and 3500 rpm for ten minutes. The supernatant was then stored at -80 °C until analysis. Cytokine concentrations were analyzed using ELISA (Meso Scale Discovery, Rockville, Maryland, USA).

*Data Analysis.* Literature shows that in humans, the cytokines take between 6 and 8 hours to return to baseline levels after the introduction of a pathogenic agent [5]. Thus, pseudodata was added at $t = 9$ and 10, to ensure that the cytokines decayed appropriately. Experimental data from all 20 subjects are shown in **Fig. 1**; pseudodata are shown in blue. The data sets marked in red represent the average dynamics of the population, and subsequent personalized simulations are shown against this data set. Box-and-whisker plots [11] were used to identify subjects displaying an abnormal response among the data sets.

*Mathematical Model*

The mathematical model developed here incorporates several key components of the acute inflammatory response, including the inflammatory trigger, here called the pathogenic ligand ($P$), resting and activated monocytes ($M_R$ and $M_A$) and circulating cytokines (TNF-$\alpha$, IL-6, CXCL8, and IL-10). These particular components were analyzed as they are regarded as main drivers of the early pro-inflammatory response (TNF-$\alpha$), the intermediate step between pro- and anti-inflammation (IL-6), neutrophil activation (CXCL8), and the late anti-inflammatory response (IL-10). The specific role of each component is described in *Table 1*. The model is formulated as a system of seven ordinary differential equations describing the dynamics of the pathogenic ligand, monocytes, and circulating cytokines and 43 parameters quantifying their interactions, illustrated in **Fig. 2.**

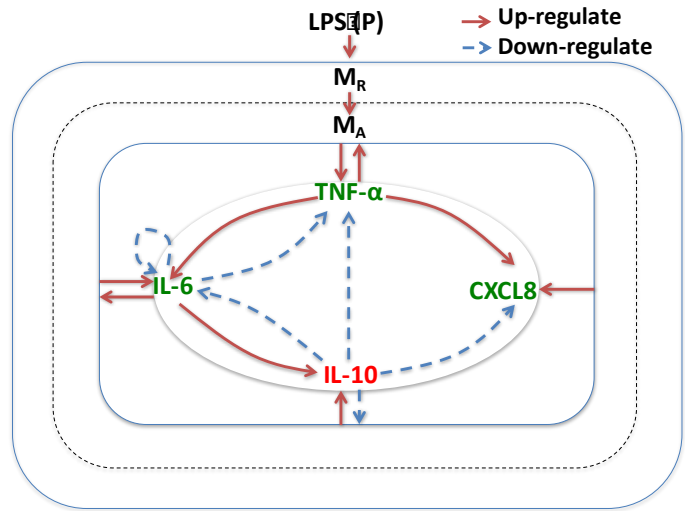

FIG 2: **Inflammatory interactions.** Intravenous injection of LPS activates circulating monocytes ($M_R$), changing them into activated monocytes ($M_A$). This begins the production of TNF-$\alpha$. At the same time, monocytes are activated to produce IL-6 and CXCL8. All three cytokines work in a positive feedback loop, amplifying the inflammatory response by activating more monocytes to stimulate production of IL-6, CXCL8, and TNF-$\alpha$. Moreover, the LPS stimulus, as well as the elevated levels of pro-inflammatory cytokines up-regulate the production of IL-10, which inhibits prolonged production of pro-inflammatory cytokines. The solid lines represent up-regulation, while the dashed line represents down-regulation.

*Pathogenic ligand.* Upon endotoxin injection, the pathogenic ligand is bound to the toll-like receptor 4 (TLR4) [12, 13] on resting circulating monocytes that will mediate its clearance from the body. In this model, the pathogenic ligand was modeled as exponentially decaying with an initial value of 2 ng/kg.

*Monocytes.* The resting monocytes are formed in the bone marrow, and are released into circulation as a means of the ongoing immune activation. The circulating monocytes are activated by endotoxin via TLR4 triggering. The circulating activated monocytes trigger the production of TNF-$\alpha$, which is responsible for the recognized signs of inflammation such as heat, increased vascular permeability and local swelling, and redness reactions [14]. The TLR4-mediated activation also leads to the production of IL-6, CXCL8, and IL-10. Moreover, several autocrine loops exist where TNF-$\alpha$ amplifies the inflammatory response by further activating monocytes to release pro-inflammatory cytokines, such as IL-6 and CXCL8. TNF-$\alpha$ also encourages the activated monocytes to produce anti-inflammatory cytokines, such as IL-10, which exert negative feedback on the system.

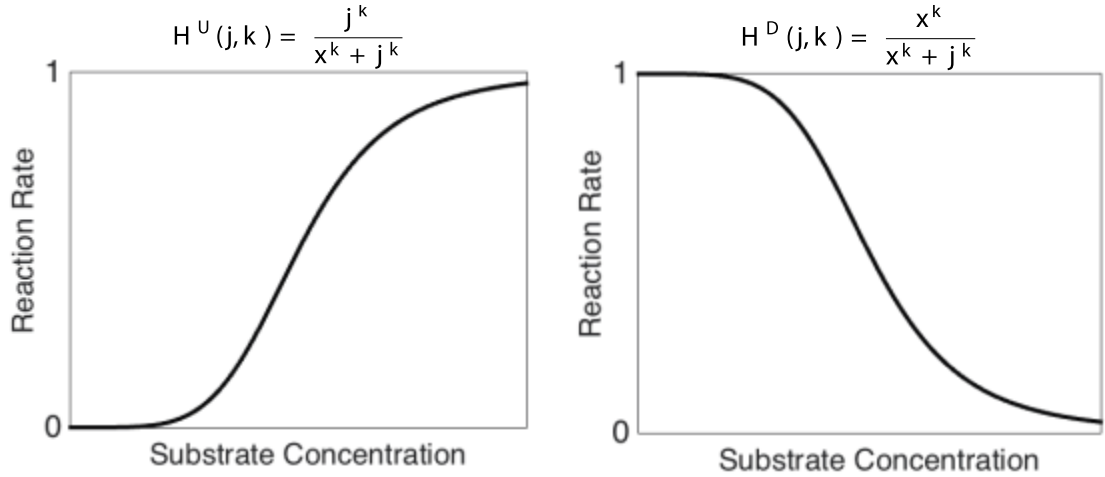

FIG 3. **Hill functions.** Up- and down-regulation functions (left and right, respectively). The reaction rate, $H$, is a function of the half-saturation value, $x$, and the substrate concentration, $j$.

The interactions between the components acting on the resting monocytes are described by the following equation

$$\dot{M}_R = -H_{up}(P)\big(k_1 + k_2 H_{up}(X)\big) H_{down}(X)\, M_R + k_3 M_R\big(1 - M_R/M_\infty\big), \qquad (1)$$

where $X \in \{\text{TNF-}\alpha, \text{IL-6}, \text{CXCL8}, \text{IL-10}\}$ and the up- and down-regulation are represented by increasing and decreasing Hill functions, $H_{up}(\cdot)$ and $H_{down}(\cdot)$, respectively; each Hill function ranges between zero and one (see **Fig. 3**). The resting monocytes are up-regulated by TNF-$\alpha$ and down-regulated by IL-10. The positive feedback of the resting monocytes on themselves is accounted for by the additional $M_R$ in the first term of the equation. The final term of **Eq. (1)** is the natural recruitment and decay of the resting monocytes modeled with a logistic growth term.

The $k_i$ terms are rate constants describing activation or elimination rates and $M_\infty$ is the maximum number of monocytes present.

The activated monocytes are represented by

$$\dot{M}_A = H_{up}(P)\left(k_1 + k_2 H_{up}(X)\right) H_{down}(X) M_R - k_4 M_A, \qquad (2)$$

which is almost identical to **Eq. (1)**, however the first term is positive and the last term is the natural decay of the activated monocytes.

*Cytokines.* The rate of change of the cytokines can be described as a combination of the number of active monocytes present and the influence from the pro- and anti-inflammatory cytokines. Mathematically, these interactions can be described by equations of the form

$$\dot{X} = \left(k_5 + k_6 H_{up}(X)\right) H_{down}(X) M_A - k_7(X - q), \qquad (3)$$

where, as before, the $k_i$ terms are rate constants describing activation or elimination rates. The down-regulation by the anti-inflammatory cytokine IL-10 (as well as IL-6, which exhibits an anti-inflammatory effect on TNF-$\alpha$ [15, 16]) is modeled by a product of decreasing Hill functions in $H_{down}$. The natural source and decay of each cytokine, that is, the behavior of the cytokine without the presence of a pathogen ligand, is represented by the last term in **Eq. (3)**. The amount of cytokine present in the absence of a pathogen is represented by the source term, $q$. For a complete list of equations, see the APPENDIX.

TABLE 1: **State variables of the mathematical model**

| Component | Function | Ref. |
| --- | --- | --- |
| Lipopolysaccharide (P) | Derived from gram-negative bacteria; induces inflammation. | [12] |
| Resting monocytes ($M_R$) | Formed in the bone marrow and are specifically transported to blood, from where they are recruited to sites of inflammation; involved in the recognition, phagocytosis, and destruction of LPS and/or pathogens. | [32, 33] |
| Activated monocyte ($M_A$) | Produces cytokines when activated by LPS. | [32, 33] |
| Tumor Necrosis Factor (TNF-$\alpha$) | Pro-inflammatory. Produced by activated monocytes and other phagocytes; amplifies inflammatory cascades; fever inducer; early. | [8, 12, 15, 32, 34] |
| Interleukin-6 (IL-6) | Pro-inflammatory. Produced by activated monocytes and other phagocytes; fever inducer; early. | [8, 12, 15, 16] |
| Chemokine Ligand 8 (CXCL8) | Pro-inflammatory. Attracts white blood cells to the site of inflammation; early. | [12, 18, 35] |
| Interleukin-10 (IL-10) | Anti-inflammatory. Limits the inflammatory response; essential for homeostasis of the immune system; late. | [8, 33, 36, 37] |

The mathematical model is comprised of the 7 ordinary differential equations. The state variables are listed in the first column. The biological implication of each component and their respective references are shown in columns 2 and 3.

*Parameterization*

To ensure that our model inputs were physiologically feasible, we studied the relationships between those used by Chow et. al, while noting the significant physiological differences between mice and humans. For instance, mice and humans differ in their cell surface recognition of LPS and downstream signal transduction [12]. These differences prevented a simple scaling to change the model from one suitable for mice to one for humans. Keeping these differences in mind, we estimated an initial parameter set.

*Initial Parameters.* The time constants of the initial parameter set were estimated to achieve model output for each cytokine and monocytes that ranged between zero and one. Then the maximum values of the human monocytes, obtained from [17], were incorporated through scaling to obtain the desired magnitude.

The half-maximum values of the Hill function were set using experimental data for the cytokines, as well as reported data in literature [5, 17-20]. In an approach similar to Clermont et. al. [9], for each equation, if cytokine $A$ was up-regulating cytokines $B$ and $C$, then the same half-maximum value was chosen to represent this interaction in the corresponding equation. That is, if the interaction between cytokines $(A \& B)$ and between $(A \& C)$ are represented by

$$\frac{A^{h_B}}{\eta_{BA}^{h_B} + A_{BA}^{h_B}} \quad \text{and} \quad \frac{A^{h_C}}{\eta_{CA}^{h_C} + A_{CA}^{h_C}},$$

respectively, then $\eta_{BA} = \eta_{CA}$. To find such a value, it was initially chosen to be approximately 60% of the max value of cytokine $A$ from the data. It was then slightly adjusted, either up or down, so that the dynamics were more in line with the data. A similar approach was used to find the half-maximum values for the down-regulatory interactions. Note that the exponents in each sigmoidal equation were able to vary between interactions; that is, $h_B \neq h_A$. These initial parameter values are shown in ***Table A2*** of the APPENDIX.

*Model Analysis*

To create a personalized model, an inverse least squares formulation was used to find a parameter set that minimized the square of the error between the computed and measured values of the cytokines. The residual vector $r$ and least squares cost $J$ are defined by

$$J = r^T r, \quad \text{where} \quad r_i = \frac{1}{\sqrt{N}} \left[ \frac{x_{\text{model}}(i) - x_{\text{data}}(i)}{x_{\text{data}}(i)} \right]^T, \tag{4}$$

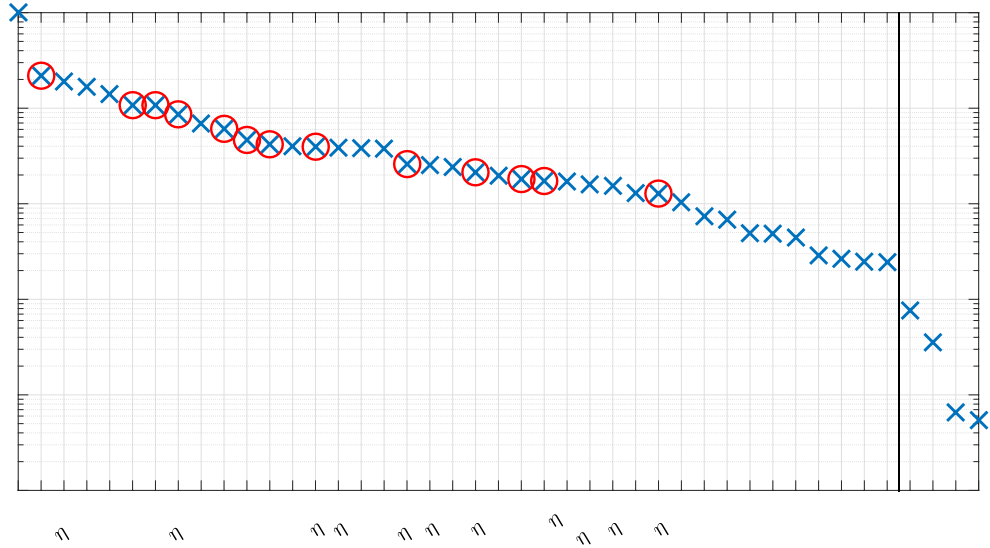

FIG 4: **Ranked Sensitivities.** Parameter sensitivities ranked from most to least sensitive. Black line shows cutoff between sensitive and insensitive parameters. Red circles denote parameters that were optimized (i.e. sensitive and identifiable).

and $N$, $x_{\text{model}}$, $x_{\text{data}}$ and $\overline{x_{\text{data}}}$ are the total number of data points, the model output, the model data, and the mean of the model data, respectively.

*Sensitivity Analysis & Subset Selection.* Sensitivity analysis separates a parameter set into the sensitive and insensitive parameters and is used to lessen the complexity of the optimization problem by reducing the parameter set. Generally, a parameter is *sensitive* if the model output is greatly affected following a slight perturbation of said parameter. A parameter is *insensitive* if the model is not affected by large perturbations in the parameter value. This can be a result of the structure of the equations as well as the nominal parameter value. A forward difference approximation, outlined in [21], was used to compute the relative sensitivities, which were then ranked from most to least sensitive. The ranked sensitivities are shown in **Fig. 4**.

Using sensitivity analysis, we concluded that 39 of the 43 parameters were sensitive. Analyzing the graphs of the Hill functions, shown in **Fig. 3**, as well as their respective equations,

$$H_{up}(X) = \frac{X^h}{\eta_{YX}^h + X^h} \quad \text{and} \quad H_{down}(X) = \frac{\eta_{YX}^h}{\eta_{YX}^h + X^h},$$

showed that for $\eta_{YX} \gg X$, $H_{up}$ approached 0, while $H_{down}$ approached 1. Conversely, for $\eta_{YX} \ll X$, $H_{up}$ approached 1, while $H_{down}$ approached 0. Without proper bounds on $\eta$, attempting to fit the model to the experimental data forced these values to become either very large or very small, depending on the equation. Finding such bounds is not trivial for these equations and our particular model. Thus, we choose to keep the half-maximum and their respective exponents fixed at their nominal values. Additionally, $M_\infty$ and the source terms $q_i$ of **Equation 3** were fixed at their nominal vales since they were set based on values found in the literature.

Based on these adjustments, our parameter set was further reduced from 39 to 14 sensitive parameters. Subset selection, via the correlation method [22], was used to separate the set into identifiable and unidentifiable parameters. A parameter is not *identifiable* if it is linearly dependent of the values of the other parameters in the model. This analysis produced a set of 13 parameters that were both sensitive and identifiable.

*Parameter Estimation.* As previously mentioned, the goal was to find a suitable parameter set that minimized the least-squares error given in **Equation 4.** The parameters were estimated using MATLAB's built-in optimization function *fminsearch,* a multidimensional unconstrained optimizer that uses the Nelder-Mead direct search method [23, 24]. Keeping the other parameters fixed, the 13 parameters identified via sensitivity analysis and subset selection were estimated. The optimal parameter values as well as their means and standard deviations are shown in *Table A3* of the APPENDIX.

*Prediction & Confidence Intervals.* After obtaining the optimized parameter values, prediction and confidence intervals were used to quantify the amount of variation in the optimized model. The *prediction interval* predicts where a single new measurement will be with a $(1-\alpha)$ probability at a given time point. It provides information about the distribution of the model output values, not the uncertainty associated with determining the population mean. To find a prediction interval, let $\hat{y}_i$ be an estimate of the model response at time $t = \hat{t}_i$, $S$ be the sensitivity matrix obtained via a forward difference approximation [21] at the times where data was collected $t = (t_0, t_1, \ldots, t_N)$, $s^2$ be an estimator of $\sigma^2$, where the estimated parameters are distributed normally with variance $\sigma^2 (S^T S)^{-1}$, and $g_i^T$ be the $i^{th}$ row of sensitivity matrix evaluated at time $t = (\hat{t}_0, \hat{t}_1 \ldots, \hat{t}_r)$. Note that, in general, $(\hat{t}_0, \hat{t}_1 \ldots, \hat{t}_r) \neq (t_0, t_1, \ldots, t_N)$; that is, the time at which the prediction interval is being calculated is different from the time at which the data was collected. The prediction interval is then given by

$$\text{PI} = \hat{y}_i \pm t_{N-M}^{\alpha/2} \, s(1 + g_i^T (S^T S)^{-1} g_i)^{1/2}, \tag{5}$$

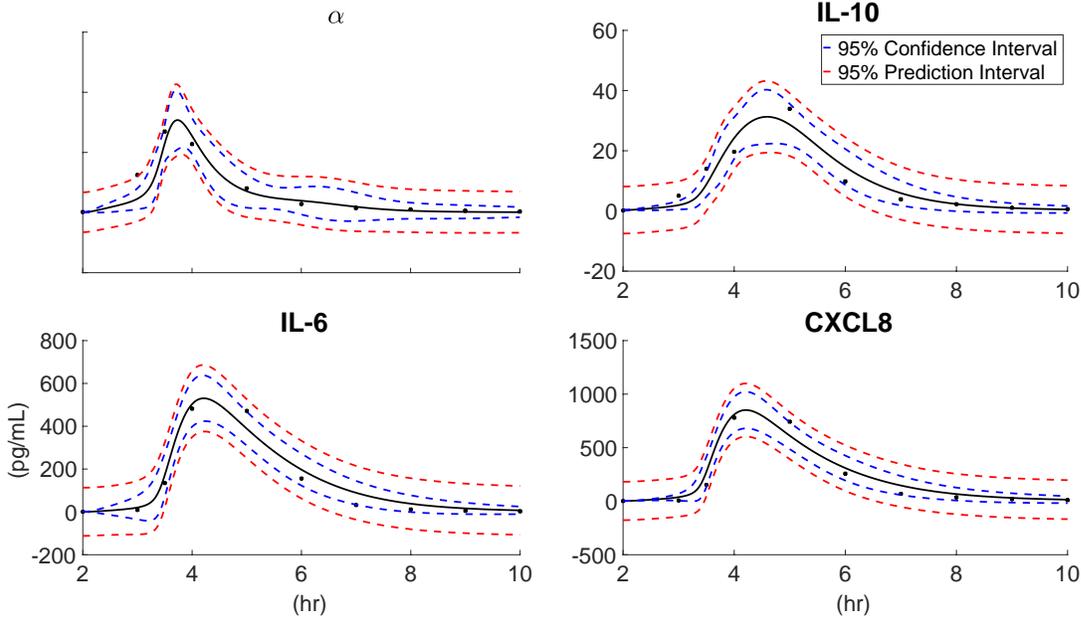

FIG 5. **Experimental data, model predictions, and prediction and confidence intervals.** Data and model predictions for one subject after intravenous administration of endotoxin (black). Confidence (blue) and prediction (red) intervals are also shown. The participant was given 2 ng/kg body weight of endotoxin at $t = 2$, and cytokine levels were measured $t = 2, 3, 3.5, 4,$ and hourly for the next 4 hours. Pseudodata was added at $t = 9$ and $t = 10$ to ensure that cytokines had appropriate time to decay.

where $N$ is the total number of data points, $M$ is the number of parameters being estimated, and $t_{N-M}^{\alpha/2}$ is the $(1 - \alpha/2)$ quantile of a Student t-distribution with $N - M$ degrees of freedom. For our analysis, we were interested in the 95% prediction interval so $\alpha = 0.05$.

The *confidence interval* measures the uncertainty of the model in predicting the mean response and is given by the following expression

$$\text{CI} = \hat{y}_i \pm t_{N-M}^{\alpha/2} \, s(g_i^T (S^T S)^{-1} g_i)^{1/2}. \qquad (6)$$

Comparing **Eqs. (5)** and **(6)**, it is noted that the prediction interval is wider than the confidence interval. The derivation of **Eqs. (5)** and **(6)** can be found in [25]. Note that in **Eq. (6)**, the second term will be 0 at $t = t_0$ since $g_i^T = 0$, due to the initial conditions. In **Fig. 1**, we highlighted the data from the subject that represented the average dynamics of the population. The model predictions along with the confidence and prediction intervals for this subject are shown in **Fig. 5.** Note that due to the nature of the equations, the prediction and confidence interval may be negative, but in practice cytokine levels are always positive.

# RESULTS

As illustrated in **Fig. 5**, the intravenous injection of LPS gives rise to rapid production of measurable cytokines in the blood stream. This response was observed to take 45 minutes to an hour to occur. Note that **Fig. 5** shows the dynamics of the cytokines after the injection of LPS. Each subject was given LPS at $t = 2$, thus the graphs also start at $t = 2$. Monocytes make up the main cell subset within the blood to respond to LPS stimulation [26]. Upon binding of LPS to TLR4, the monocytes will become activated and respond with production of cytokines where some will be secreted rapidly after activation, while others require *de novo* biosynthesis, and will have a longer response time. The circulating activated monocytes induce the production of TNF-$\alpha$, at 1.5 to 2 hours after the initial injection of LPS. The activated monocytes also begin production of IL-6 and CXCL8 at 2 to 2.5 hours after the injection. The anti-inflammatory cytokine IL-10 makes a later appearance [12] than the pro-inflammatory cytokines, being measurable at 2.75 to 3 hours upon LPS exposure. This production inhibits continuous inflammation. Evidence of this is shown in the fact that the time at which the pro-inflammatory cytokines began to decrease coincides with the time that IL-10 peaked, around 2.25 to 2.75 hours. Once the LPS is cleared from the bloodstream, the cytokines will return to their baseline levels.

The optimized parameter values are shown in **Table A3**. We used the prediction and confidence intervals shown in **Fig. 5** to quantify the accuracy of the model and the parameters ability to depict the data. Individual response data that discern from the mean data set are identified as dashed lines in **Fig. 1.** The model prediction for an individual showing an abnormal response along with the mean data set is shown in **Fig. 6.**

## *Statistical Assessment*

The $R^2$ statistic was used to quantify the ability of the model to accurately depict the data. When fitting the model to the mean of the available data, we obtained $R^2$ values of 0.91, 0.9, 0.97, and 0.97 for TNF-$\alpha$, IL-10, IL-6, and CXCL8 respectively. Thus, we can conclude that the model presents an overall good fit to the data. The prediction and confidence intervals shown in **Fig. 5** show that the model accurately depicts the mean response and that the optimized parameter values provide a reasonable prediction of the data.

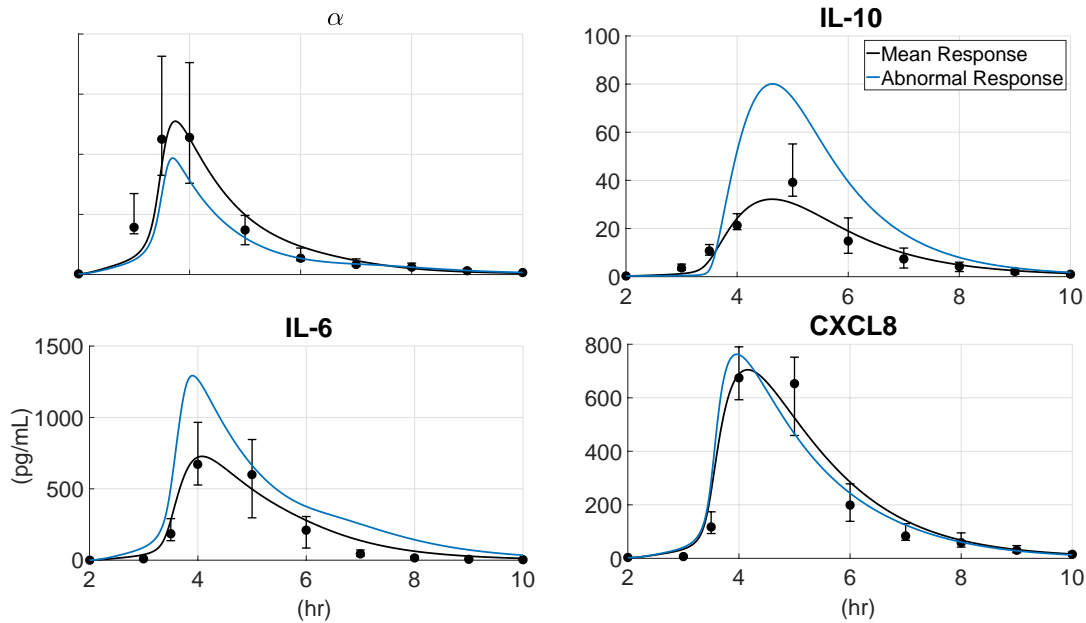

FIG 6. **Model predictions for the mean data set and an abnormal individual response data set.** Mean data set (black circles), model predictions for the mean data set (black line), interquartile range (error bars), and model predictions for an abnormal data set (blue line), are shown. Participants were given 2 ng/kg body weight of endotoxin at $t = 2$, and cytokine levels were measured $t = 2, 3, 3.5, 4$, and hourly for the next 4 hours. Pseudodata was added at $t = 9$ and $t = 10$ to ensure that cytokines had appropriate time to decay.

## DISCUSSION

In this study we generated a personalized cytokine response model based on systemic measures of cytokine production after low dose i.v. endotoxin injection in healthy men. The specific constraints related to individual response levels made it necessary to build a personalized model. Biological variations as the one seen in the data set shown in **Fig. 6** is a common phenomenon in individuals responding to LPS, and therefore it is of great importance to be able to model all individual response types. It is clear that the response in the abnormal subject is significantly different than that of the mean data set for TNF-$\alpha$, IL-10, and IL-6. This could be due to the fact that IL-6 has an anti-inflammatory effect on TNF-$\alpha$ [2, 16], so a high level of IL-6 can result in lower levels of TNF-$\alpha$. Additionally, TNF-$\alpha$ and IL-10 have opposing roles in the inflammatory response, which can lead to a high level of IL-10 when TNF-$\alpha$ is low. Such differences in cytokine response patterns in monocytes to LPS stimulation have been studied at the molecular level in *ex vivo* stimulation assays. Based on these studies, it appears that individual differences in the expression and regulation of the interferon regulatory factor 3 pathway may play a role in regulating the level of specific cytokine production after LPS activation [27].

These differences in the responses also caused noteworthy contrasts in the parameter values. For instance, $k_{6M}$, the rate responsible for the amount of up-regulation of IL-6 from the activated monocytes, was much larger for abnormal response data (mean sets: 0.293, abnormal set: 1.33). As expected, this resulted in much higher levels for IL-6. Conversely, $k_{6TNF}$, the rate responsible for the amount of TNF-$\alpha$ secreted from activated monocytes, was lower in the abnormal response set (mean sets: 1, abnormal set: 0.342). This can explain the lower levels of TNF-$\alpha$ measured in this individual.

One of the limitations of the model was the sparseness of the available data points. The majority of the changes in the dynamics of the system occur between 1 and 3 hours after the administration of LPS. Thus, having more data points available during those times, perhaps every 15 minutes, may have provided more accurate results. For example, we expect TNF-$\alpha$ to peak 1.5 to 2 hours after the subject is given LPS. However, because the data was collected with 30 min intervals, at times t = 1, 1.5, 2 hours we cannot predict exactly when the cytokine level peaks. Additionally, having data for the numbers of circulating resting and activated monocytes may have provided insight as to why a particular subject has a specific response to the endotoxin versus another since the monocytes are responsible for the measured cytokine cascade. Another limitation was estimating the parameters in the Hill functions. To make the model personalized, we may require different half-maximum values and exponents for each subject. However, we were unable to find an effective way to estimate them without having the Hill functions operate on the tail end (close to either one or zero).

Besides the modeling of direct triggering of IL-10 production after TLR4 ligation by LPS, previous models [8] have also included an interaction between IL-10 and TNF-$\alpha$ in which IL-10 is up-regulated by TNF-$\alpha$. However, biological evidence supporting this claim has not been found, which leads us to believe that it might not be a direct interaction. Instead, it could be that the up-regulation of IL-6 from TNF-$\alpha$ [15, 28] induces an increase in the levels of IL-10. To investigate this, sensitivity analysis, subset selection, and optimization were performed on a new model containing the interaction between IL-10 and TNF-$\alpha$. Once a suitable parameter set and model fit were obtained, the Akaike Information Criterion (AIC), a model selection tool used to compare different models quantitatively [29], was used to measure the goodness of fit with and without that interaction. AIC measures the amount of information lost when a given model is used to describe the behavior of a system, so the smaller the value the better. **Table A4** shows the calculated AIC values for the model with and without IL-10 being up-regulated by TNF-$\alpha$. The AIC values along with each model's prediction of the data, led to the conclusion that it would be best to omit this hypothetical/possible pathway.

Although we chose to study TNF-$\alpha$, IL-6, CXCL8, and IL-10, there are many other important cytokines and factors involved in the inflammatory process. For example, IL-1$\beta$ is considered one of the most important pro-inflammatory cytokines released from monocytes upon LPS-induced activation [12] and nitric oxide promotes inflammation and tissue injury [8, 14]. Additionally, the endotoxin-signaling pathway has been shown to involve lipopolysaccharide-binding protein (LBP), and the co-activators myeloid differentiation-2 (MD-2), and CD14, as well as TLR4. Each signaling pathway has a specific reaction time that may be dependent on the dose of endotoxin, the availability of co-activators as well as the specific cytokine being activated. For instance, Blomkalns et al. [30] found that the release of CXCL8 by freshly isolated human PBMCs given a low dose of endotoxin was dependent on both membrane-associated CD14 and TLR4. In addition, it has been found that recognition by a specific receptor cluster is associated with the strain of bacteria [31]. To increase the accuracy of our model, we need to identify the specific pathways activated and the time necessary for production of the specific cytokine, and then incorporate these factors into our model.

In conclusion, we have developed a personalized mathematical model of the inflammatory response in humans. This model was built on sequential measures of cytokines in circulating blood, but might be connected to highly correlated parameters that are more easily monitored in patients during early postoperative mobilization. Despite having sparse data and difficulties in estimating certain parameters, we were able to make a personalized model. In addition to cytokine levels, we also have blood pressure and heart rate data available from this experiment. We hope to expand the current model to include blood pressure and heart rate, which can provide an idea of why some patients faint after surgery while others do not. This may lead to the development of preoperative therapy that can be used to shorten a patient's hospital stay, reducing heath care costs and improving patient's quality of life.

# APPENDIX

## Model Equations

Below are the differential equations modeling the dynamics of the inflammatory response; they represent the interactions depicted in **Fig. 1**. Each up- or down-regulation is represented by a sigmoidal function of the form

$$H_Y^U(X) = \frac{X^h}{\eta_{YX}^h + X^h} \quad \text{or} \quad H_Y^D(X) = \frac{\eta_{YX}^h}{\eta_{YX}^h + X^h},$$

respectively. Here, $X$ is the cytokine (or pathogenic ligand, in the case of the monocytes) responsible for either up- or down-regulating component $Y$. The half-maximum value is represented by $\eta_{YX}$ and the associated exponent is $h$. These functions are non-dimensional and range between zero and one. The nominal parameter values and units are shown in **Table A2.**

TABLE A1: **State variables and equations of the mathematical model**

| | |
|---|---|
| Pathogen | $\dfrac{dP}{dt} = -k_P P$ |
| Resting Monocytes | $\dfrac{dM_R}{dt} = -H_M^U(P)\left(k_M + k_{MTNF} H_M^U(TNF)\right) H_M^D(IL10) M_R + k_{MR} M_R \left(1 - M_R/M_\infty\right)$ |
| Activated Monocytes | $\dfrac{dM_A}{dt} = H_M^U(P)\left(k_M + k_{MTNF} H_M^U(TNF)\right) H_M^D(IL10) M_R - k_{MA} M_A$ |
| Interleukin-6 | $\dfrac{dIL6}{dt} = \left(k_{6M} + k_{6TNF} H_{IL6}^U(TNF)\right) H_{IL6}^D(IL6) H_{IL6}^D(IL10) M_A - k_6 (IL6 - q_{IL6})$ |
| Tumor Necrosis Factor | $\dfrac{dTNF}{dt} = k_{TNFM} H_{TNF}^D(IL6) H_{TNF}^D(IL10) M_A - k_{TNF} (TNF - q_{TNF})$ |
| CXCL8 | $\dfrac{dIL8}{dt} = (k_{8M} + k_{8TNF} H_{IL8}^U(TNF)) H_{IL8}^D(IL10) M_A - k_8 (IL8 - q_{IL8})$ |
| Interleukin-10 | $\dfrac{dIL10}{dt} = \left(k_{10M} + k_{106} H_{IL10}^U(IL6)\right) M_A - k_{10} (IL10 - q_{IL10}).$ |

TABLE A2: **Nominal parameter values and units**

| No. | Parameter | Value | Unit | No. | Parameter | Value | Unit |
|---|---|---|---|---|---|---|---|
| 1 | $k_{10}$ | 0.8 | $hr^{-1}$ | 23 | $h_{106}$ | 3.68 | - |
| 2 | $k_{10M}$ | 0.0191 | $\frac{pg}{mL \cdot hr \cdot \# \, of \, cells}$ | 24 | $h_{6TNF}$ | 2 | - |
| 3 | $k_6$ | 0.66 | $hr^{-1}$ | 25 | $h_{66}$ | 1 | - |
| 4 | $k_{6M}$ | 0.81 | $\frac{pg}{mL \cdot hr \cdot \# \, of \, cells}$ | 26 | $h_{610}$ | 4 | - |
| 5 | $k_8$ | 0.66 | $hr^{-1}$ | 27 | $h_{8TNF}$ | 3 | - |
| 6 | $k_{8M}$ | 0.56 | $\frac{pg}{mL \cdot hr \cdot \# \, of \, cells}$ | 28 | $h_{810}$ | 1.5 | - |
| 7 | $k_{TNF}$ | 1 | $hr^{-1}$ | 29 | $h_{TNF10}$ | 3 | - |
| 8 | $k_{TNFM}$ | 0.6 | $\frac{pg}{mL \cdot hr \cdot \# \, of \, cells}$ | 30 | $h_{TNF6}$ | 2 | - |
| 9 | $k_{MA}$ | 2.51 | $hr^{-1}$ | 31 | $h_{M10}$ | 0.3 | - |
| 10 | $k_{MR}$ | 0.006 | $hr^{-1}$ | 32 | $h_{MTNF}$ | 3.16 | - |
| 11 | $k_P$ | 1.01 | $hr^{-1}$ | 33 | $h_{MP}$ | 1 | - |
| 12 | $\eta_{610}$ | 34.8 | $pg/mL$ | 34 | $q_{TNF}$ | 1.08 | $pg/mL$ |
| 13 | $\eta_{66}$ | 560 | $pg/mL$ | 35 | $q_{IL10}$ | 0.248 | $pg/mL$ |
| 14 | $\eta_{6TNF}$ | 185 | $pg/mL$ | 36 | $q_{IL8}$ | 1.42 | $pg/mL$ |
| 15 | $\eta_{810}$ | 17.4 | $pg/mL$ | 37 | $q_{IL6}$ | 0.317 | $pg/mL$ |
| 16 | $\eta_{8TNF}$ | 185 | $pg/mL$ | 38 | $k_M$ | 0.0414 | $hr^{-1}$ |
| 17 | $\eta_{106}$ | 560 | $pg/mL$ | 39 | $M_\infty$ | 30000 | # of cells |
| 18 | $\eta_{TNF10}$ | 17.4 | $pg/mL$ | 40 | $k_{6TNF}$ | 0.81 | $\frac{pg}{mL \cdot hr \cdot \# \, of \, cells}$ |
| 19 | $\eta_{TNF6}$ | 560 | $pg/mL$ | 41 | $k_{8TNF}$ | 0.56 | $\frac{pg}{mL \cdot hr \cdot \# \, of \, cells}$ |
| 20 | $\eta_{MP}$ | 3.3 | $ng/kg$ | 42 | $k_{106}$ | 0.0191 | $\frac{pg}{mL \cdot hr \cdot \# \, of \, cells}$ |
| 21 | $\eta_{M10}$ | 4.35 | $pg/mL$ | 43 | $k_{MTNF}$ | 8.65 | $hr^{-1}$ |
| 22 | $\eta_{MTNF}$ | 100 | $pg/mL$ | | | | |

TABLE A3: **Optimal parameter values.**

| Parameter | Nominal Value | Optimized Value | Mean ±Standard Deviation |
|---|---|---|---|
| $k_{10M}$ | 0.019 | 0.028 | 0.017 ± 0.011 |
| $k_{10}$ | 0.8 | 1.10 | 0.899 ± 0.230 |
| $k_6$ | 0.66 | 0.903 | 0.947 ± 0.315 |
| $k_{6M}$ | 0.81 | 0.295 | 0.481 ± 0.282 |
| $k_8$ | 0.66 | 0.857 | 0.883 ± 0.258 |
| $k_{TNF}$ | 1 | 2.22 | 1.74 ± 0.524 |
| $k_{TNFM}$ | 0.6 | 1 | 0.798 ± 0.276 |
| $k_{MA}$ | 2.51 | 2.32 | 2.88 ± 1.37 |
| $k_P$ | 1.01 | 0.631 | 0.641 ± 0.331 |
| $k_{6TNF}$ | 0.81 | 1 | 1.18 ± 0.581 |
| $k_{8TNF}$ | 0.56 | 1.50 | 0.830 ± 0.697 |
| $k_{106}$ | 0.0191 | 3.65e-4 | 0.012 ± 0.020 |
| $k_{MTNF}$ | 8.65 | 2.97 | 7.69 ± 9.97 |

Optimal parameter values and mean plus/minus standard deviation excluding abnormal data sets shown in **Figure 1**.

TABLE A4. **Akaike Information Criterion Results.**

|  | without interaction | with interaction |
|---|---|---|
| **TNF-$\alpha$** | **99.435** | 108.977 |
| **IL-6** | **105.887** | 114.929 |
| **IL-8** | **110.812** | 118.333 |
| **IL-10** | **77.648** | 87.4316 |

AIC values for model without TNF-$\alpha$ up-regulating IL-10 (column 2), and a model with the interaction (column 3). Values shown in bold indicate lowest value for particular state variable and model.

# FIGURE LEGENDS

FIG 1. **Experimental Data.** Plasma cytokine responses to intravenous endotoxin administration in 20 healthy young men. Median (black circle), interquartile range (error bars), and subject most in line with data mean (red) are depicted. Abnormal response (identified via Box-and-Whisker plots) are denoted by dashed lines. Pro-inflammatory cytokines, TNF-$\alpha$, IL-6, and CXCL8, and the anti-inflammatory cytokine IL-10 levels were measured at t = 0, 2, 3, 3.5, 4 h and in one hour increments for the next 4 hours. Pseudodata was added at $t = 9$ and 10 to ensure that cytokines decayed to baseline levels (blue). Endotoxin was administered at t = 2.

FIG 2: **Inflammatory interactions.** Intravenous injection of LPS activates circulating monocytes ($M_R$), changing them into activated monocytes ($M_A$). This begins the production of TNF-$\alpha$. At the same time, monocytes are activated to produce IL-6 and CXCL8. All three cytokines work in a positive feedback loop, amplifying the inflammatory response by activating more monocytes to stimulate production of IL-6, CXCL8, and TNF- $\alpha$. Moreover, the LPS stimulus, as well as the elevated levels of pro-inflammatory cytokines up-regulate the production of IL-10, which inhibits prolonged production of pro-inflammatory cytokines. The solid lines represent up-regulation, while the dashed line represents down-regulation.

FIG 3. **Hill functions.** Up- and down-regulation functions (left and right, respectively). The reaction rate, $H$, is a function of the half-saturation value, $x$, and the substrate concentration, $j$.

FIG 4: **Ranked Sensitivities.** Parameter sensitivities ranked from most to least sensitive. Black line shows cutoff between sensitive and insensitive parameters. Red circles denote parameters that were optimized (i.e. sensitive and identifiable).

FIG 5. **Experimental data, model predictions, and prediction and confidence intervals.** Data and model predictions for one subject after intravenous administration of endotoxin (black). Confidence (blue) and prediction (red) intervals are also shown. The participant was given 2 ng/kg body weight of endotoxin at $t = 2$, and cytokine levels were measured $t = 2, 3, 3.5, 4$, and hourly for the next 4 hours. Pseudodata was added at $t = 9$ and $t = 10$ to ensure that cytokines had appropriate time to decay.

FIG 6. **Model predictions for the mean data set and an abnormal individual response data set.** Mean data set (black circles), model predictions for the mean data set (black line), interquartile range (error bars), and model predictions for an abnormal data set (blue line), are shown. Participants were given 2 ng/kg body weight of endotoxin at $t = 2$, and cytokine levels were measured $t = 2, 3, 3.5, 4$, and hourly for the next 4 hours. Pseudodata was added at $t = 9$ and $t = 10$ to ensure that cytokines had appropriate time to decay.